\documentclass[prl,aps,twocolumn,floatfix]{revtex4}
\usepackage{graphics}
\begin{document}
\title{Exotic $p$-wave superfluidity of single hyperfine state Fermi gases in optical lattices}
\author{M. Iskin and C. A. R. S{\'a} de Melo}
\affiliation{School of Physics, Georgia Institute of Technology, Atlanta, Georgia 30332, USA}
\date{\today}

\begin{abstract}
We consider $p$-wave (triplet) pairing of single hyperfine state ultracold atomic gases trapped 
in quasi-two-dimensional optical lattices.  
We find that the critical temperatures in the lattice model is considerably higher and experimentally 
attainable around half-filling in contrast to the predictions of continuum model for $p$-wave superfluids.
In tetragonal lattices, we show that the atomic compressibility and spin susceptibility
have a peak at low temperatures exactly at the half-filling, but this peak splits into two 
in the orthorhombic lattices. These peaks reflect the $p$-wave structure of the order parameter 
for superfluidity and they disappear as the critical temperature is approached from below. 
We also calculate the superfluid density tensor, and show that for the orthorhombic case there is no
off-diagonal component, however in the tetragonal case an off-diagonal component develops, 
and becomes a key signature of the exotic $p$-wave state. 

\pacs{PACS:}PACS: 03.75.Ss, 03.75.Hh
\end{abstract}
\maketitle

Tunable optical lattices have been extensively used to study phase transitions 
in bosonic atomic gases~\cite{opt4,greiner2}, 
because they allow the controlled manipulation of the particle density $n$, 
and of the ratio between the particle transfer matrix element $t$, and the interparticle interaction 
strength $V$~\cite{regal1,ohara}. 
This kind of control is not fully present in standard fermionic condensed 
matter systems, and has hindered the development of experiments that could probe systematically 
the effects of strong correlations as a function of $n$ and $t/V$. However, fermionic atomic gases like
$^6{\rm Li}$ and $^{40}{\rm K}$ have been succesfully trapped,
and their normal state and superfluid properties are beginning to be studied~\cite{modugno,kohl,jaksch,hofsfetter,torma1}. 
Since superfluid phases are more easily accessible 
in the experiments involving ultracold atomic gases, 
spin-polarized ultracold atomic systems 
are ideal candidates for the observation of novel triplet superfluid phases
and for testing theoretical models that were proposed earlier.
Thus, it is only natural to propose that optical lattices could be used to study the normal state
and superfluid properties of ultracold fermionic systems as a function of $n$, $t/V$ and
lattice symmetry. This systems are of a broad interest not only for the
atomic physics community but also for the nuclear, condensed matter 
and more generally for the many-body physics communities
where superfluidity models have been investigated in various contexts.

Presently there is only experimental evidence that 
$^{40} {\rm K}$ (Ref.~\cite{regal2,greiner}) and $^6{\rm Li}$ (Ref.~\cite{hulet,litium1,litium2,litium3,kinast}) 
can form weakly and tightly bound 
atom pairs, when the magnetic field is swept through an $s$-wave Feshbach resonance.
However, the properties of $p$-wave spin-polarized ultracold fermions 
and their possible superfluid behaviour are beginning to be investigated~\cite{john,regal3}.
When identical fermionic atoms are trapped in a single hyperfine state (SHS) the interaction between them is strongly 
influenced by the Pauli exclusion principle, which prohibits $s$-wave scattering of atoms in identical spin states.
As a result, in SHS degenerate Fermi gases, two fermions 
can interact with each other at best via $p$-wave scattering. 
Thus, one expects that the superfluid
ground state of such SHS Fermi gases to be $p$-wave and spin triplet. 

In the $p$-wave channel, if the atom-atom interactions are effectively attractive then 
the onset for the formation of Cooper pairs in three dimensions occurs at a temperature  
$T_c\approx E_F\exp[-\pi/(2k_F a_{sc})^3]$ in the BCS limit,
where $E_F$ is the Fermi energy and $a_{sc}$ is the $p$-wave scattering length. 
Unfortunately, this temperature is too low to be observed experimentally. However, 
in the presence of the Feshbach resonances~\cite{john,regal3}, $p$-wave interactions can be enhanced, and
the critical temperature for superfluid is expected to increase to experimentally accessible values. 
On the other hand, we show that the spin triplet ($p$-wave) weak coupling limit in optical lattices 
(like in the singlet cases~\cite{hofsfetter}) is sufficient to produce a superfluid critical temperature 
that is accessible experimentally.

In this manuscript, we consider quasi-two-dimensional optical lattices
with a periodic trapping potential of the form $U(r)=\sum_{i} U_{0,i}\cos^2(k_ix_i)$,
with $U_{0,z}\gg \min \{U_{0,x},U_{0,y} \}$, which strongly suppresses tunneling along the $z$ direction. This is 
a non-essential assumption, which just simplifies the calculations, but still describes an experimentally relevant
situation. Here  $x_i = x, y,~{\rm or}~ z$ labels the spatial coordinates, $k_i=2\pi/\lambda_i$ is the wavelength, 
and $U_{0,i}$  is the potential well depth along direction $x_i$, respectively.
The parameters $U_{0,i}$ are proportional to the laser intensity along each direction, and it is typically several 
times the one photon recoil energy $E_R$ such that tunneling is small and the tight-binding approximation can be used.

Thus, in the presence of magnetic field $\mathbf{h}$, we consider the following quasi-two-dimensional lattice 
Hamiltonian (already in momentum space) for an SHS Fermi gas
\begin{eqnarray}
\label{eqn:hamiltonian}
H=\sum_{\mathbf{k}}\xi(\mathbf{k})a_{\mathbf{k}\uparrow}^\dagger a_{\mathbf{k}\uparrow} + 
\frac{1}{2}\sum_{\mathbf{k},\mathbf{k'},\mathbf{q}}V(\mathbf{k},\mathbf{k'}) b_{\mathbf{k},\mathbf{q}}^\dagger b_{\mathbf{k'},\mathbf{q}}, 
\end{eqnarray}
where the pseudo-spin $\uparrow$ labels the trapped hyperfine state represented by 
the creation operator $ a_{\mathbf{k}\uparrow}^\dagger$, and
$b_{\mathbf{k},\mathbf{q}}^\dagger=a_{\mathbf{k}+\mathbf{q}/2,\uparrow}^\dagger a_{-\mathbf{k}+\mathbf{q}/2,\uparrow}^\dagger$.
Furthermore, 
$\xi(\mathbf{k})=\varepsilon(\mathbf{k})-\tilde{\mu}$ describes the tight-binding dispersion 
$\varepsilon(\mathbf{k}) = - t_x \cos k_xa_x - t_y \cos k_ya_y - t_z \cos k_za_z $, 
with $\tilde{\mu}=\mu+g\mu_Bh$, and $\min \{t_x, t_y\} \gg t_z$. 
$V(\mathbf{k},\mathbf{k'})=w^\dagger(\mathbf{k})\mathbf{V}w(\mathbf{k'})$ 
is the $p$-wave pairing interactions with 
matrix elements $V_{ij} = -2V_{0,i}\delta_{ij}$ and 
$w^\dagger(\mathbf{k})=(\sin k_xa_x, \sin k_ya_y)$.
Here, $V_{0,i}> 0$ is the effective interaction and $a_i$ is the corresponding lattice length along
the $i^{th}$ direction.

Using the functional integration formalism~\cite{popov} ($\beta=1/T$ and units $\hbar=k_B=1$), 
we obtain the saddle point effective action
\begin{eqnarray}
\frac{S_0}{\beta}=\Delta_0^\dagger\frac{\mathbf{V}^{-1}}{2} \Delta_0
+ \sum_{p}\left(\frac{\xi(\mathbf{k})}{2}-\frac{1}{\beta}\rm{Tr}\ln\frac{\mathbf{G}^{-1}(p)}{2}\right), 
\end{eqnarray}
where we use $p=(\mathbf{p},iv_l)$ with $v_l=(2l + 1 )\pi/\beta$ and define 
stationary vector field $\Delta_0^\dagger=(\Delta_{0,x}^*, \Delta_{0,y}^*)$.
Here $\mathbf{G}^{-1}(p)/\beta=iv_\ell \sigma_0-\xi(\mathbf{k})\sigma_3 + 
\Delta_0^\dagger w(\mathbf{k})\sigma_- + w^\dagger(\mathbf{k}) \Delta_0 \sigma_+$ 
is the inverse Nambu propagator and $\sigma_{\pm}=(\sigma_1\pm\sigma_2)/2$ and $\sigma_{i}$ is the Pauli spin matrix.
The condition $\delta S_0 /\delta \Delta_0^* = 0$ leads to the order parameter equation
\begin{equation}
\Delta_0 = \mathbf{M}\Delta_0 \label{gap}
\end{equation}
where $\mathbf{M}$ has matrix elements 
$M_{ij}=\sum_{\mathbf{k}}V_{0,i}\sin k_ia_i\sin k_ja_j \tanh(\beta E(\mathbf{k})/2) / E(\mathbf{k})$. 
Here, $E(\mathbf{k})=(\xi^2(\mathbf{k})+|\Delta(\mathbf{k})|^2)^{\frac{1}{2}}$ is the quasi-particle energy and 
the scalar order parameter $\Delta(\mathbf{k})= w^\dagger(\mathbf{k})\Delta_0$ is separable in temperature $T$ and momentum $\mathbf{k}$. 

Within the irreducible representations of the D$_{4h}$ (D$_{2h}$) group in the tetragonal (orthorhombic) lattices,
~\cite{annett} our exotic $p$-wave state corresponds to the $^3E_u(n)$ representation with a $d$-vector given by 
\begin{equation}
d(\mathbf{k})=f(\mathbf{k})(1,i,0) \label{dvector}, 
\end{equation}
where $f(\mathbf{k})=AX+BY$, and $X$ and $Y$ 
are $\sin k_xa_x$ and $\sin k_ya_y$, respectively. Notice that, 
this state also breaks time reversal symmetry,
as expected from a fully spin-polarized state.
In the tetragonal lattice, the stable solution for our model corresponds to the case $A=B\ne0$, and thus
to the $^3E_u(d)$ representation,  
where spin-orbit symmetry is preserved, but both spin and orbit symmetries are independently 
broken. In the orthorhombic lattice, the stable solutions correspond to either $A\ne0,B=0$
or $A=0, B\ne0$, thus leading to the $^3E_u(b)$ representation. 

Our main interest is in tetragonal (square) lattices, however, we also want to investigate the effects of 
small anisotropies in optical lattice lengths. 
For definiteness, we set $a_y = a$ constant and investigate square and orthorhombic lattices 
where $a_x = a$ and $a_x = a (1 - \delta)$ with $\delta \ll 1$, respectively.
In the case of square lattices, we choose the parameters 
$t_x= t_y = t$ and $V_{0,x}=V_{0,y} = V_0 = 0.3t$ 
and change $V_{0,x}$ and $t_x$ accordingly as we vary $a_x$.
Using exponentially decaying on-site Wannier functions and the WKB approximation, 
we obtain that the tunneling and interaction matrix elements along the $x$-direction are 
proportional to $\exp{(2\delta\sqrt{U_{0,x}/E_R})}$.
%%, where $U_{0,x}$ is the amplitude of the optical potential along the $x$-direction and $E_R$ is the one photon recoil energy. 
In the present calculation, we take $t_x=t\exp{(10\delta)}$ and $V_{0,x}=V_0\exp{(10\delta)}$.
As we increase (decrease) the ratio $\delta$, both interactions and tunneling rate increase (decrease) in $\hat{x}$
direction. A different choice of on-site Wannier functions does not change our general 
conclusions, the only qualitative difference is that $t_x$ and $V_{0,x}$ are different functions of $\delta$. 
Depending on the lattice anisotropy, there are three distinct solutions: 
(1) $\Delta_{0,x}\ne0, \Delta_{0,y}=0$ ($A \ne 0, B = 0$);
(2) $\Delta_{0,x}=0, \Delta_{0,y}\ne0$ ($ A = 0, B \ne 0$); and (3) $\Delta_{0,x}= \Delta_{0,y}\ne 0$ ($A = B \ne 0$).
In a square lattice both directions are degenerate (case 3), but even a small anisotropy in the lattice 
spacings or in the lattice potential lifts the degeneracy and throws the system into either case (1) or (2).

The critical temperature, $T_c = \max\{ T_{c,x}, T_{c,y} \}$, is determined from the condition $\det\,\mathbf{M}=1$ in Eq.~(\ref{gap}), and can be written as
\begin{eqnarray}
0=\prod_{i=x,y}\left( 1-V_{0,i}\sum_{\mathbf{k}}\frac{\sin ^2k_ia_i}{\xi(\mathbf{k})}
\tanh \frac{\xi(\mathbf{k})}{2T_{c,i}} \right) \label{tc}.
\end{eqnarray}
But, this equation has to be solved simultaneously with the saddle point number equation 
$N=-\partial \Omega_0/\partial \tilde{\mu}$ where $\Omega_0=S_0(\Delta(\mathbf{k}))/\beta$ is the saddle point 
thermodynamic potential. This leads to 
$N=\sum_{\mathbf{k}}n(\mathbf{k})$ 
where 
\begin{equation}
n(\mathbf{k})=\frac{1}{2}\left[ 1 - \frac{\xi(\mathbf{k})}{E(\mathbf{k})}\tanh\frac{\beta E(\mathbf{k})}{2} \right]
\end{equation}
is the momentum distribution. These equations are approximately valid for all temperatures $(T \le T_c)$ 
in the limit of weak interactions (BCS limit). 
\begin{figure} [ht]
\centerline{\scalebox{0.37}{\includegraphics{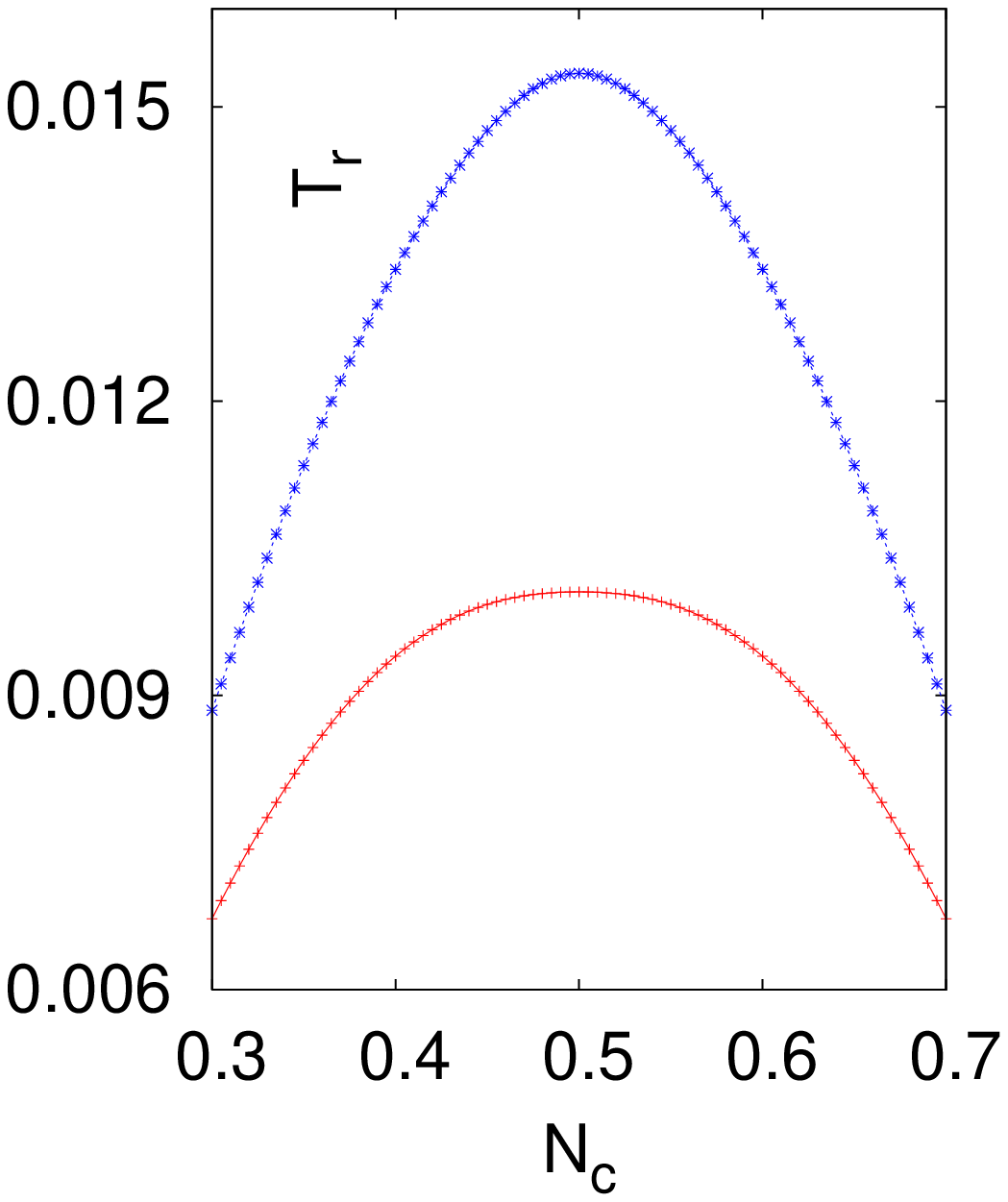}}
\scalebox{0.37}{\includegraphics{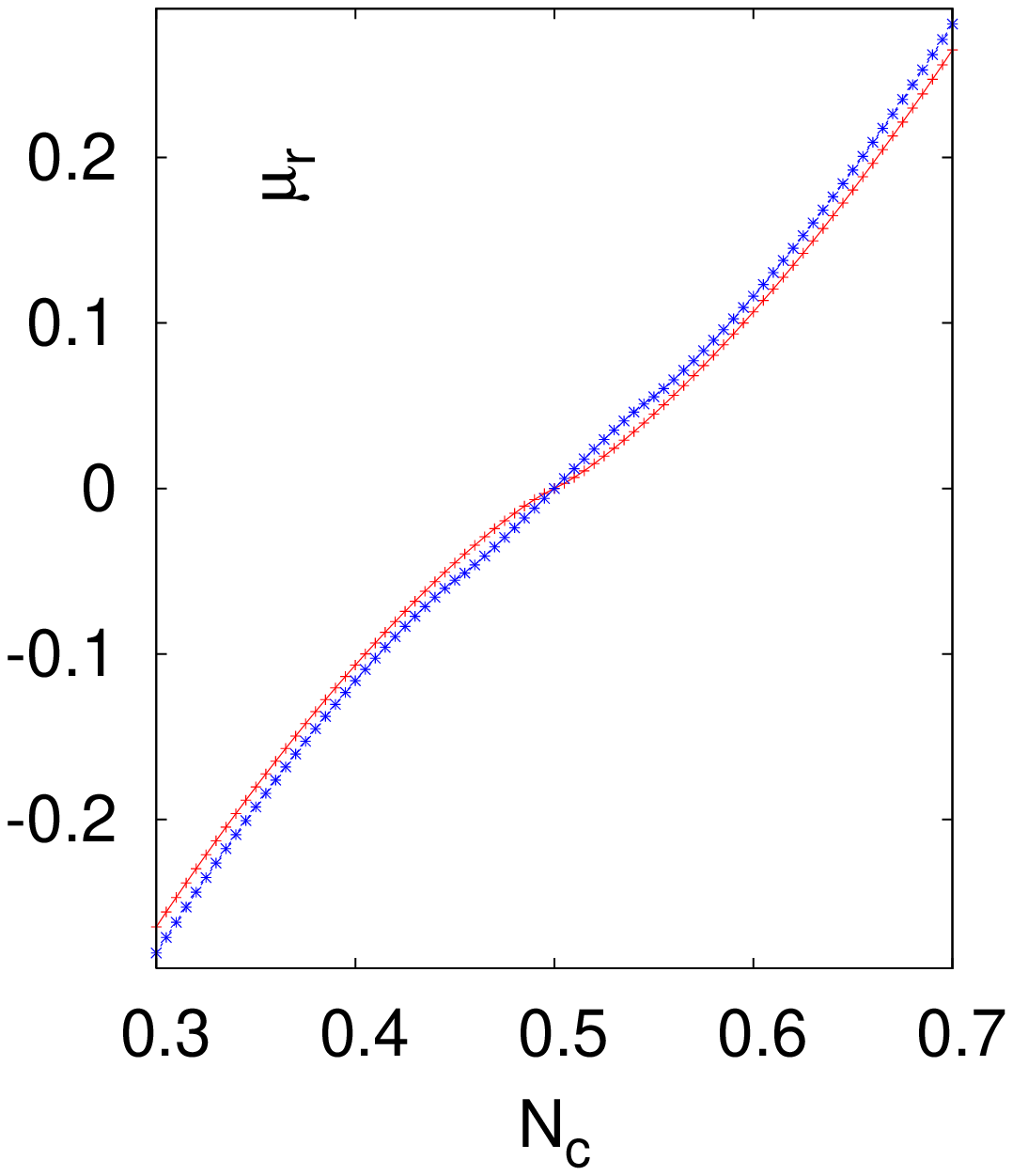}}
}
\caption{\label{ngtc} (Colour online) Plot of a) critical temperature $T_r = T_c/E_0$, and b) chemical potential $\mu_r = \tilde{\mu}/E_0$ 
versus filling $N_c$ for the square case $a_x = a$ (red); and the orthorhombic case $a_x = 0.99a$ (blue).
}
\end{figure}
Plots of $T_r = T_c/E_0$ and $\mu_r = \mu/E_0$ as a function of number of 
atoms per unit cell ($N_c$) are shown in Fig.~1 for the cases of square
and orthorhombic lattices. 
Here, $E_0 = 2t$ is the half-filling Fermi energy for the square lattice.  
Notice that $T_r$ is maximal at half-filling, and that 
it has a value $0.01$, which is much higher than the theoretically predicted $T_c$ from the continuum model, 
and comparable to experimentally attainable $T/T_F \approx 0.01$. This implies that the superfluid regime of 
spin-polarized fermion gases may be observed experimentally in a lattice, 
even in the limit of weak interactions. The observability of a superfluid transition in spin-polarized
fermion systems is clearly enhanced when the system is driven through a Feshbach resonance (in a lattice or in the
continuum), as $T_c$ is expected to increase further in this case, however our calculations indicate 
that the weak interaction (BCS) limit may be sufficient in the lattice case.

Notice that the order parameter symmetry dramatically effects $T_c$ 
as can be seen by rewriting Eq.~(\ref{tc}) as
\begin{equation}
1 = \prod_{i=x,y}V_{0,i} \int_{-t_x-t}^{t_x+t} d\varepsilon 
\frac{\tanh\frac{\varepsilon - \mu}{2T_{c,i}}}{2(\varepsilon - \mu)}
D_{p,i}(\varepsilon),
\end{equation}
where we define an effective density of states (EDOS)
$
D_{p,i}(\varepsilon) = \sum_{\mathbf{k}} \delta[\varepsilon - \varepsilon(\mathbf{k})](\sqrt{2}\sin k_ia_i)^2
$
which is plotted in Fig.~2 for $a_x = a$ (black), $a_x = 0.99a$ (red), and $a_x = 0.95a$ (blue).
Here, $\sqrt{2}\sin k_ia_i$ are symmetry factors related to the order parameter.
For instance in the $s$-wave case, this symmetry factor is 1 and EDOS becomes
$
D(\varepsilon) = \sum_{\mathbf{k}} \delta[\varepsilon - \varepsilon(\mathbf{k})],
$ 
which is the density of states (DOS) of normal fermions. 
\begin{figure} [ht]
\centerline{\scalebox{0.36}{\includegraphics{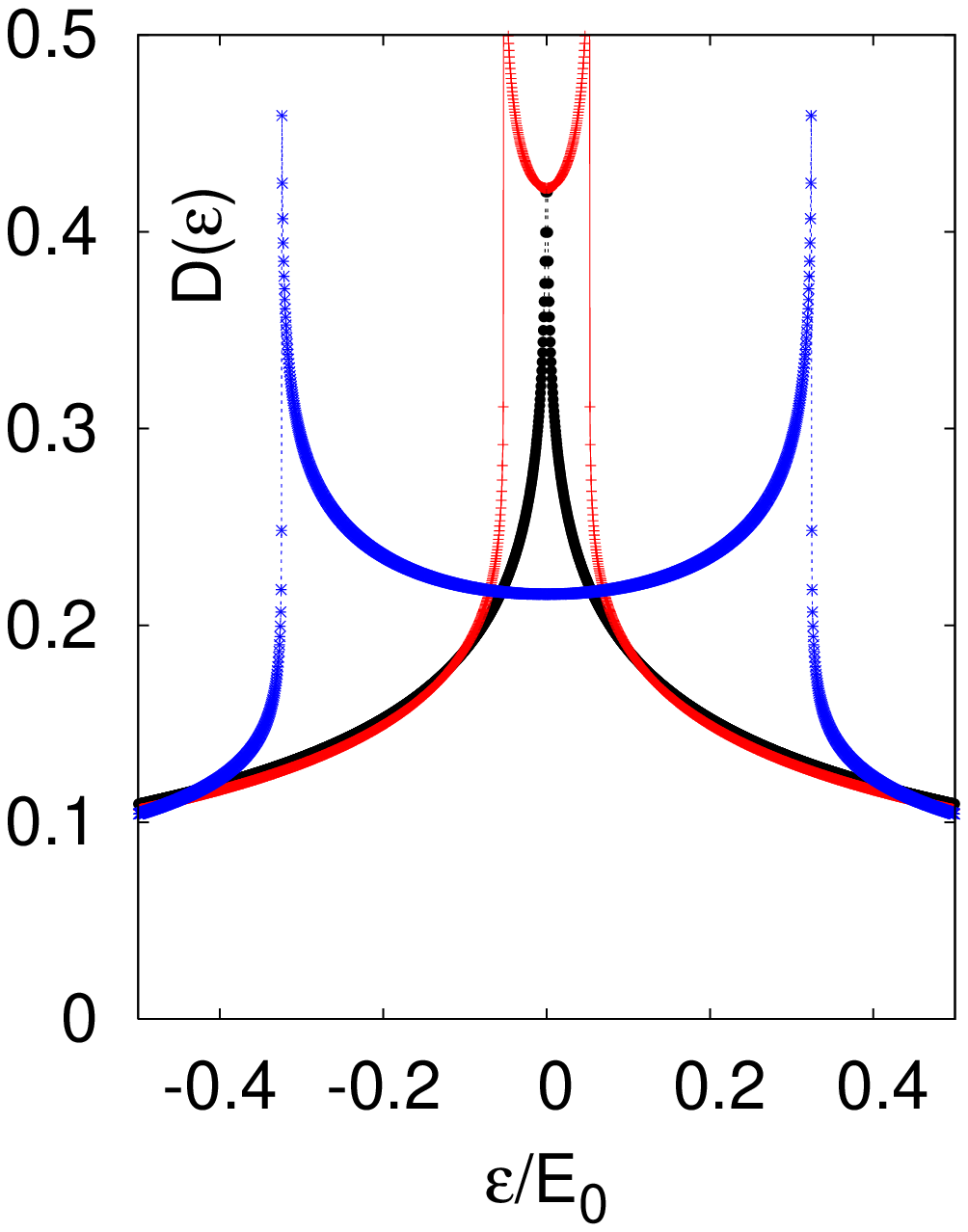}}
\scalebox{0.36}{\includegraphics{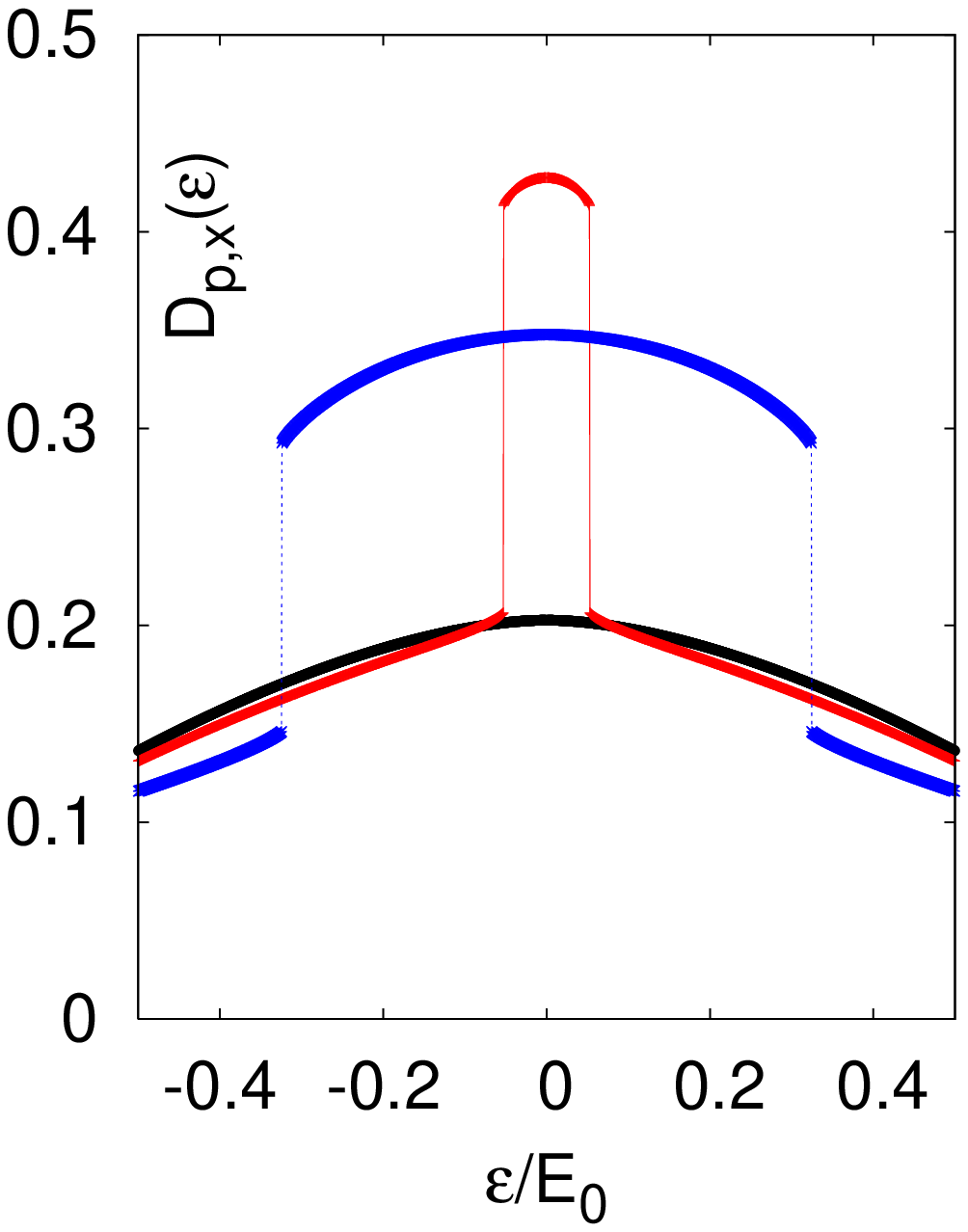}}
}
\caption{\label{dos} (Colour online) Plot of a) normal state DOS $D(\varepsilon)$ and 
b) $p$-wave EDOS $D_{p,x}(\varepsilon)$ versus energy $\varepsilon / E_0$
in units of $t$ for $a_x = a$ (black) and $a_x = 0.99a$ (red), and $a_x = 0.95a$ (blue)
in two dimensions.
}
\end{figure}
For the $p$-wave symmetry discussed, $D_{p,i}(\varepsilon) V_{0,i}$ 
plays the role of a dimensionless coupling parameter which controls the critical temperature. 
Notice that, EDOS and $T_c$ are maximal at half-filling in square lattices ($a_x = a$). 
Additionally, EDOS decreases and finally vanishes at the band edges 
where a small ratio of $T_c/T_F$ was predicted from continuum models. 
However, this is not the case around half-filling and we expect weak interactions 
to be sufficient in observing superfluidity.
In the orthorhombic lattices, EDOS and $T_c$ are considerably increased and are
maximal around half-filling with a small anisotropy ($a_x = 0.99a$) in the lattice spacings.
Notice, however, that further anisotropy in the lattice spacings ($a_x = 0.95a$)
leads to a decrease in EDOS and $T_c$ around half-filling with respect to the case with $a_x = 0.99a$. 

Next we discuss the atomic compressibility and the spin susceptibility of the system.
The isothermal atomic compressibility is given by 
$ \kappa=-(1/N_c^2)\partial^2\Omega / \partial \tilde{\mu}^2
=(1/N_c^2)\partial N_c / \partial \tilde{\mu}$
where
\begin{equation}
\label{dndmu}
\frac{\partial N_c}{\partial \tilde{\mu}} = 
\sum_{\mathbf{k}} \frac{\Delta^2(\mathbf{k})}{2E^3(\mathbf{k})} \tanh \frac{\beta E(\mathbf{k})}{2} 
+ \sum_{\mathbf{k}} Y(\mathbf{k})\frac{\xi^2(\mathbf{k})}{E^2(\mathbf{k})} 
\end{equation}
and $Y(\mathbf{k})=(\beta/4)\rm{sech}^2[\beta E(\mathbf{k})/2]$. 
Plots of $\kappa_r = \kappa/\kappa_0$, where normalization $\kappa_0$ is evaluated at $T=0$ and half-filling, 
is shown in Fig.~3 for the square and orthorhombic lattices. 
In the square lattice case, $\kappa$ has a peak at half-filling
and low temperatures, and a hump at $T_c$. The same 
qualitative behavior is obtained in the case of orthorhombic lattices, 
with the additional feature that the central peak (hump) splits into two due to degeneracy lifting. 
\begin{figure} [ht]
\centerline{\scalebox{0.36}{\includegraphics{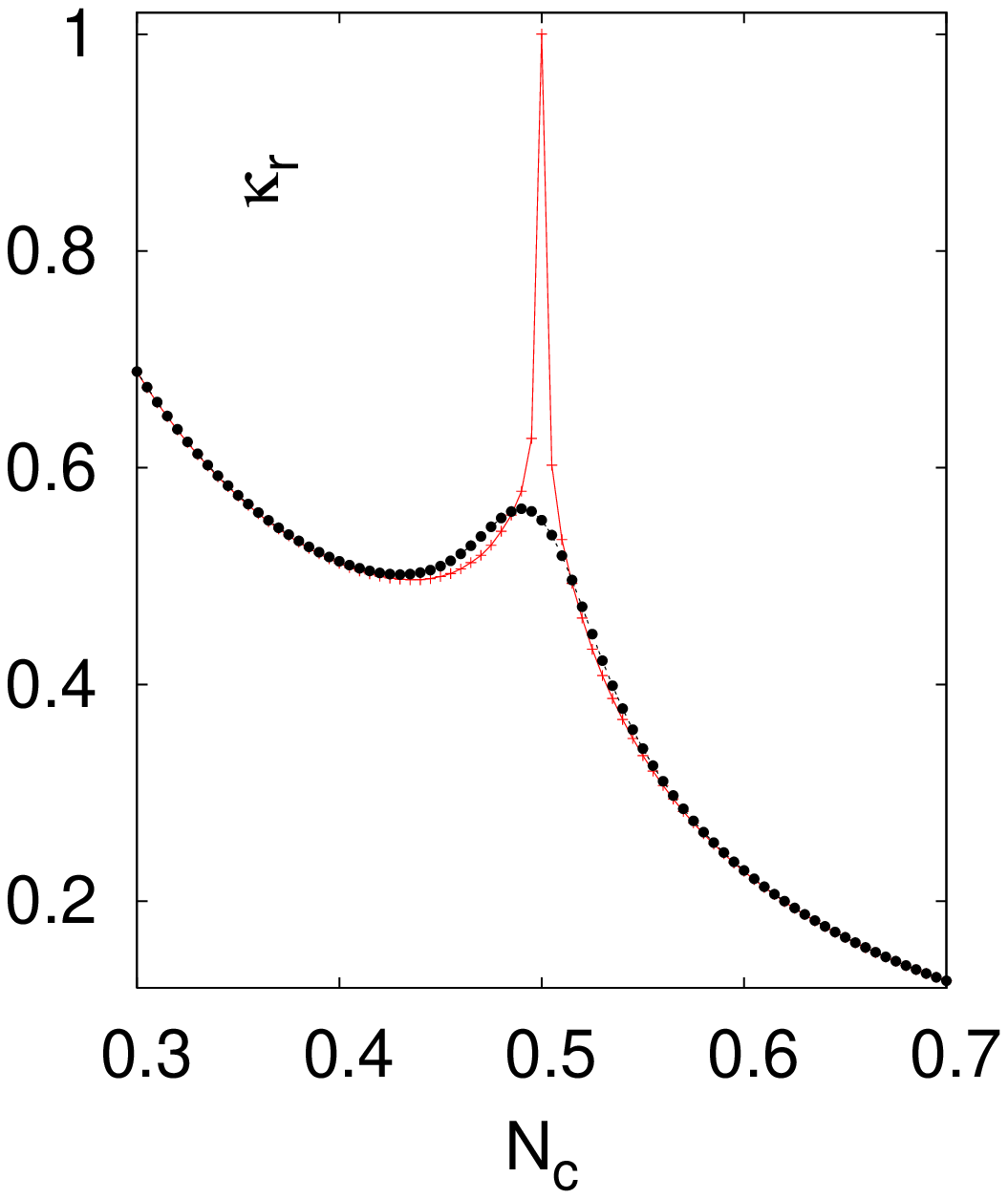}}
\scalebox{0.36}{\includegraphics{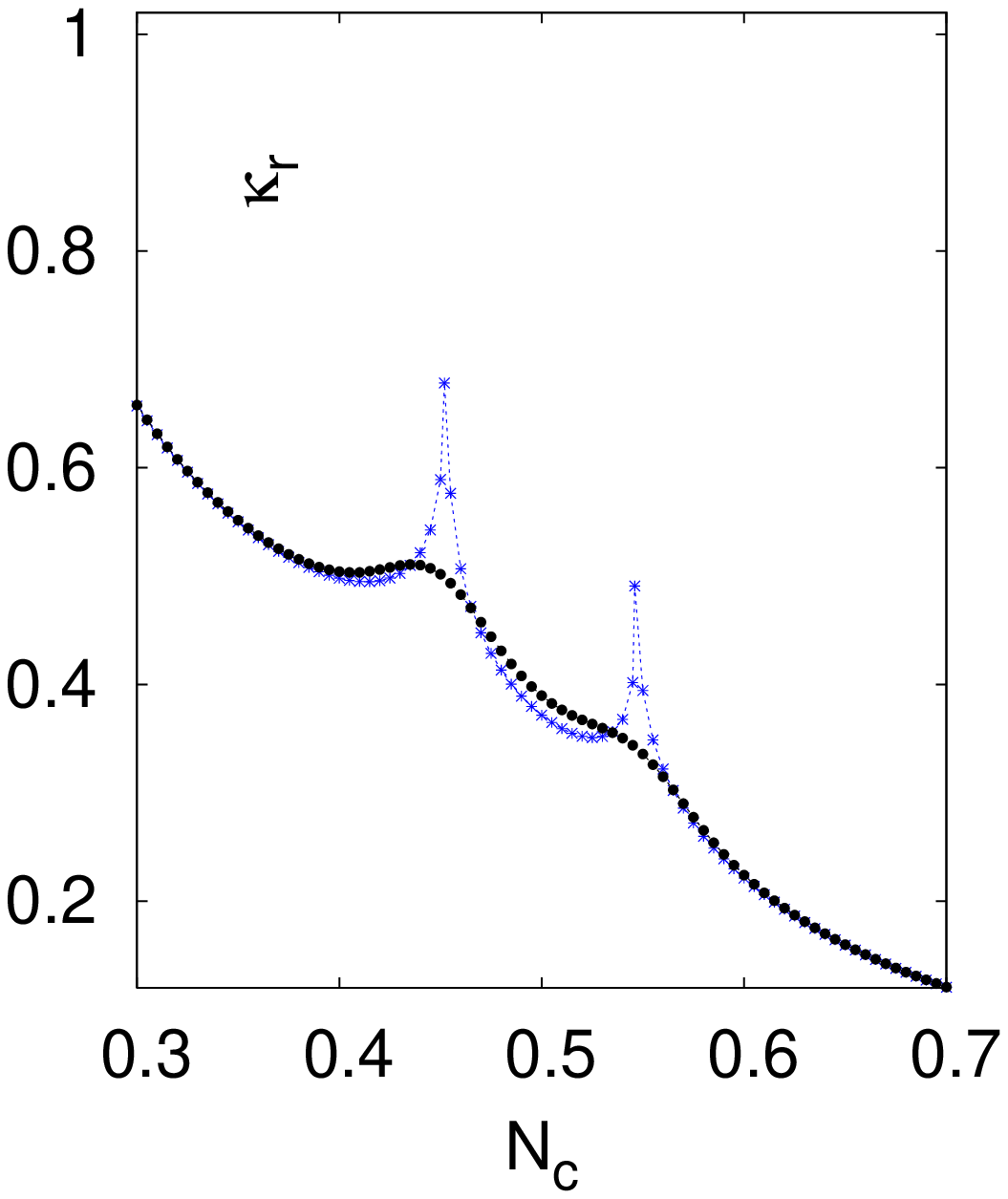}}
}
\caption{\label{ngtc} (Colour online) Plot of compressibility $\kappa_r = \kappa/\kappa_0$ versus
filling $N_c$ at $T=0.004T_c^{max}$ for a) $a_x = a$ (red) and b) $a_x = 0.99a$ (blue). 
Notice that the compressibility peaks disappears at $T=T_c$ and turns into humps (black) in both cases.
}
\end{figure}

The peaks at $T \approx 0$, can be understood by noting that 
$N_c^2 \kappa$ can be written as $\sum_{\mathbf{k}} g(\mathbf{k}) / E(\mathbf{k})$
for $T = 0$, where $g(\mathbf{k}) = 2n(\mathbf{k}) [1 - n(\mathbf{k})]$.
Therefore, the peaks are due to non-vanishing $g(\mathbf{k})$
in regions of $\mathbf{k}$-space where $E(\mathbf{k})$ vanishes and
$n(\mathbf{k})$ is rapidly changing. 
In tetragonal lattices, the integrand $g(\mathbf{k})/ E(\mathbf{k})$ 
has 4 $\mathbf{k}$-space points ($0, \pm \pi$), and ($\pm \pi, 0$)
in the first Brillouin zone (1BZ) where it diverges only 
when the chemical potential $\mu = 0$ ($N_c = 0.5$). 
Similarly in orthorhombic lattices, the integrand diverges  
at 2 $\mathbf{k}$-space points ($\pm \pi, 0$) in the 1BZ 
when $\mu = t_x - t = 0.021$ ($N_c = 0.548$).
Furthermore,  the integrand diverges at 2 $\mathbf{k}$-space points ($0, \pm \pi$) when
$\mu = -t_x + t = -0.021$ ($N_c = 0.452$). 
For every other $\mu$ the integrand is well-behaved resulting in a 
smooth $\kappa$ in both cases.
At $T = T_c$, the humps are not related to the order parameter, but are due 
to the peaks appearing in DOS (Fig. 2a).
Notice that, while DOS have only one divergence at half-filling in the tetragonal case,
it has two peaks in the case of orthorhombic lattices which leads to two humps.  

Furthermore, for a magnetic field ${\bf h} = h {\bf \hat \eta}$ applied along
an arbitrary $\eta$-direction, the spin susceptibility tensor component is
$\chi_{\eta\eta}=-\partial^2 \Omega / \partial h^2
= g^2\mu_B^2 \partial N_c / \partial \tilde{\mu}$. 
Therefore, in spin-polarized systems, $\chi_{\eta \eta}$ is directly related to the atomic compressibility
and is given by $[N_c^2/(g\mu_B)^2]\kappa$.

Next we discuss the superfluid density of the system. 
Generally speaking, there are two components to the reduction of the superfluid density at non-zero 
temperatures, one coming from fermionic (quasi-particle excitations) and
the other bosonic (collective modes) degrees of freedom. 
The fermionic component comes from phase twists of the order parameter 
and the temperature dependence of its components is given by
\begin{equation}
\rho_{ij}=\frac{1}{2V}\sum_{\mathbf{k}}\left[ n(\mathbf{k})\partial_{i}\partial_{j}\xi(\mathbf{k}) 
-Y(\mathbf{k})\partial_{i}\xi(\mathbf{k})\partial_{j}\xi(\mathbf{k})\right],
\end{equation}
where $n(\mathbf{k})$ is the momentum distribution, and $\partial_i$ denotes the partial derivative with respect to $k_i$.
Here, we will not discuss the bosonic contribution, except to say that 
at low temperatures the dominant terms come from Goldstone modes 
associated with the phase of the order parameter, which 
are underdamped in our case due to sub-critical Landau damping. 
Furthermore, in the case of tetragonal symmetry,
Goldstone modes do not contribute to the off-diagonal component of the superfluid density, 
which is the main focus of the analysis that follows.
\begin{figure} [ht]
\centerline{\scalebox{0.55}{\includegraphics{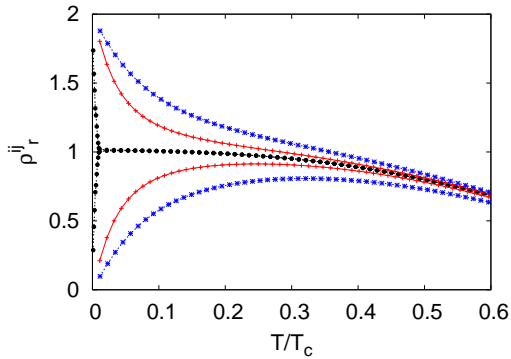}}}
\caption{\label{ngsdvsn} (Colour online) Plot of diagonal ($\rho^{xx}_r = \rho^{yy}_r$; upper curves) 
and off-diagonal ($\rho^{xy}_r = \rho^{yx}_r$; lower curves) 
superfluid density tensor components $\rho^{ij}_r = \rho_{ij} / \rho_{xy}^{max}$
versus temperature $T/T_c$ at fillings $N_c=0.5$ (black), $0.45$ (red) and $0.4$ (blue) when $a_x = a$.
}
\end{figure}

In Fig.~4, we plot $\rho^{ij}_r = \rho_{ij} / \rho_{xy}^{max}$
as a function of temperature for three fillings; $N_c = 0.5$ (black), $N_c = 0.45$ (red), and $N_c = 0.4$ (blue).
Normalization $\rho_{xy}^{max}$ is the maximum value of the off-diagonal component 
and is evaluated at half-filling and $T=0.02T_c$ in a square lattice.
It is important to emphasize that square lattices have identical diagonal elements $\rho_{xx}=\rho_{yy}$ 
due to the tetragonal symmetry, but have a nonzero $\rho_{xy}$ component (Fig.~4) as a result of the absence of reflection 
symmetry in the $yz$-plane ($x \to -x$) or the $xz$-plane ($y \to -y$) for the $d$-vector defined in Eq.~(\ref{dvector}). 
Notice that, reflection symmetry is restored and 
$\rho_{xy}$ vanishes identically in the orthorhombic case for any temperature $T$.
We show that while $\rho_{xy}$ is zero at $T = 0$ for any filling,
it increases until all tensor components have the same value at some higher temperature.

In summary, we considered $p$-wave pairing of single-hyperfine-state Fermi gases in quasi-two-dimensional
optical lattices. 
We found that the critical temperatures in tetragonal and orthorhombic optical lattices
are considerably higher than the continuum model predictions, and therefore, experimentally 
attainable.
At low temperatures, we found a peak in the atomic compressibility 
(and similarly in the spin susceptibility) exactly at half-filling for the tetragonal lattice.
This peak splits into two smaller peaks in the orthorhombic case.
These peaks reflect the $p$-wave structure of the order parameter at low temperatures, 
and they decrease in size as
the critical temperature is approached from below. 
Furthermore, in the orthorhombic lattices, 
the off-diagonal component $\rho_{xy}$ of the superfluid density tensor vanishes identically, while
the diagonal components $\rho_{xx}$ and $\rho_{yy}$ are different.  
However, in the square lattices, we showed that $\rho_{xy} \ne 0$, while $\rho_{xx} = \rho_{yy}$. 
The presence of non-zero $\rho_{xy}$ is a key signature of our exotic $p$-wave triplet state. 

{\it Acknowledgement}: We would like to thank NSF (DMR-0304380) for financial support.

\end{document}